**DeepBeat: A multi-task deep learning approach to assess signal quality and arrhythmia detection in wearable devices**

Authors: Jessica Torres Soto[1], Euan A. Ashley[2]*

[1] Department of Biomedical Informatics, Stanford University, Stanford, CA, USA
[2] Department of Medicine, Division of Cardiovascular Medicine, Stanford University, Stanford, CA, USA

# Abstract

Wearable devices enable theoretically continuous, longitudinal monitoring of physiological measurements like step count, energy expenditure, and heart rate. Although the classification of abnormal cardiac rhythms such as atrial fibrillation from wearable devices has great potential, commercial algorithms remain proprietary and tend to focus on heart rate variability derived from green spectrum LED sensors placed on the wrist where noise remains an unsolved problem. Here, we develop a multi-task deep learning method to assess signal quality and arrhythmia event detection in wearable photoplethysmography devices for real-time detection of atrial fibrillation. We train our algorithm on over one million simulated unlabeled physiological signals and fine-tune on a curated dataset of over 500K labeled signals from over 100 individuals from 3 different wearable devices. We demonstrate that, in comparison with a traditional random forest based approach (AF events, precision:0.24, recall:0.58, f1:0.34, auPRC:0.44) and a single task CNN (AF events, precision:0.59, recall:0.69, f1:0.64, auPRC:0.68) our architecture using unsupervised transfer learning through convolutional denoising autoencoders dramatically improves the performance of atrial fibrillation detection in participants at rest (AF events, pr:0.94, rc:0.98, f1:0.96, auPRC:0.96). Finally, we validate algorithm performance on a prospectively derived replication cohort of ambulatory subjects using data derived from an independently engineered device. We show that two-stage training can help address the unbalanced data problem common to biomedical applications, where large-scale well-annotated datasets are hard to generate due to the expense of manual annotation, data acquisition, and subject privacy. Further, our algorithm addresses noise—an essential challenge in wearable technology applications. In conclusion, though a combination of simulation and transfer learning and we develop and apply a multitask architecture, to the problem of atrial fibrillation detection from wearable wrist sensors demonstrating high levels of accuracy and a solution for the vexing challenge of mechanical noise.

# Introduction

Wearable devices are increasingly used in cardiology for out-of-the-clinic healthcare monitoring[1]. Such platforms enable physiological parameters like ECG, heart rate, heart rhythm, and physical activity to be measured with greater frequency, convenience and improved accuracy [2,3]. The fast expansion of these wearable device functionalities in a healthcare setting can help engage individuals in understanding disease progression and allow for the detection of early disease trajectories[1]. Wrist-based smartwatch sensing for healthcare has generally been focused on photoplethysmography (PPG) for the detection of heart rate and heart rhythm. Of the abnormal heart rhythms, atrial fibrillation (AF), characterized by a disorganized chaotic electrical activity of the atria, has received the most attention. The incentive for studying AF detection methods in wearable devices is two-fold. First, AF can often go unnoticed and yet is a risk factor for stroke: early AF detection should allow simple interventions that could greatly decrease stroke risk. Secondly, within the next 25 years, as a result of the increased aging population in industrialized nations, the prevalence of AF is expected to double. There will be an increasing public health need for cost-effective methods for AF detection and monitoring.

Current wearable based algorithms aimed at AF detection use heart rate variability metrics over extended time periods. Some more recent papers have adopted deep learning methods[3]. These methods commonly remain proprietary or lack sufficient information or data for easy reproducibility [4–9]. Noise in wrist-worn wearable devices remains an unsolved problem[10]. Inaccurate heart rate estimation and misdetection of AF are largely dependent on poor signal quality [11,12]. Correctly estimating and assessing signal input quality is critical for accurate AF detection methods. Historically event detection for abnormal cardiac events relied on explicit rules and domain expert knowledge to craft features with high discriminatory power. Noise artifact detection methods mirrored this approach as well, with calculating feature-based approaches to estimate signal quality[13]. Published methods include Root Mean Square of the Successive Difference of peak-to-peak intervals (RMSSD), Shannon entropy (ShE), Poincaré plot analysis (PPA), dynamic time warping for shape analysis, and spectral analysis [14–16]. However, these methods generally rely on distance-based metrics, which in many situations, have been shown to yield unreliable results [11].

PPG signals are vulnerable to many types of artifacts including motion artifacts, improper wear, skin tone, and presence of tattoos. Recent work in developing an optimal signal quality index (SQI) relies on manually selected features to try to distinguish high-quality PPG signals from poor or signals [6,17]. The limitations of manually selected features include the challenge of designing consistent descriptors/features for diverse PPG environments, across different individuals, while maintaining high discriminative power [17]. Our method for signal quality assessment does not rely on descriptors/features but instead on learned features, that are generally found to outperform manually selected ones in the case of heterogeneous and unfamiliar datasets [18].

Recently, convolutional neural networks (CNN), a class of artificial neural networks, with a strong capability in feature extraction, have achieved great success in computer vision medical tasks [19–21]. Features are no longer hand-designed but, rather, learned by models trained through back-propagation [21]. CNN's have become the dominant choice for many machine learning tasks due to their high discriminatory power in supervised machine learning settings, a facet that relies on large quantities of manually labeled data for building high quality models. There are, however, limitations to CNNs, including their sensitivity to weight initialization and their dependency on large-scale labeled training data. In some domains, like biomedical applications, it is very difficult to construct large-scale well-annotated datasets due to the cost of manual annotation, data acquisition, and patient privacy. This can limit development and accessibility. Thus, the ability to learn effectively from smaller datasets or unlabeled observations is critical to alleviating the bottlenecks in the advancement of this paradigm for improving healthcare.

Transfer learning aims to solve the problem of insufficient training data. In transfer learning, the goal is to build models that adapt and generalize well even when data distributions between datasets used for training and testing differ.

Some researchers have applied deep convolutional neural networks to the problem of AF event detection [4–9]. *Shashikumar et al* [4] developed a blended approach, combining the output of a CNN with other selected features derived from beat-to-beat variability and signal quality. This method expanded upon the feature extraction process with the help of feature engineering by experts to define appropriate and critical features needed for success. *Tison et al.* [5] proposed using averaged heart rates, step count and time-lapse as input for passive detection of AF using a neural network consisting of 8 layers, each of which had 128 hidden units. Lastly, *Poh et al.* [6] proposed a method for dense convolutional neural networks to distinguish between noise, sinus rhythm, ectopic rhythms, and AF across an ensemble of 3 simultaneously collected PPG signals. However, a joint task of AF classification and quality assessment calls has not yet been addressed nor any exploration of transfer learning to boost discriminatory power has been considered. Providing a quality assessment score with each rhythm classification call allows for high-quality scores to signify that a rhythm call is more reliable and less likely to be incorrect. In addition, exploring transfer learning for AF detection is attractive given limited access to large labeled cohorts that are common in biomedical research.

To address this gap, we present DeepBeat, a method for the detection of atrial fibrillation from wrist-based PPG sensing. Our method combats the unique noise artifact problem common in AF detection by utilizing a multi-task convolutional neural network architecture, transfer learning (TL) and an auxiliary signal quality estimation task, for AF event detection from spatially segmented physiological PPG signals. We investigate the use of unsupervised transfer learning from convolutional denoising auto-encoders to show that unsupervised pre-training improves the performance tasks of physiological signal quality assessment and AF event detection.

# Results

*Training the model*

The training was broken into two phases: pre-training using CDAE on over one million simulated physiological signals and fine-tuning using transfer learning on a collected set of real-world data. The real-world dataset composed of data collected at Stanford University from subjects undergoing elective cardioversions or elective stress tests and supplemented with a publicly available dataset from the IEEE Signal Processing Cup 2015 (Table 1)**.** The data was split into training, validation, and test set with no subject overlap between sets. We evaluate the model's performance on two datasets, the first is the held-out test set and the second is an external evaluation dataset collected from a different commercial device that is composed of 15 individuals monitored in an ambulatory approach. The test dataset is reflective of our training set in terms of data acquisition in a controlled environment, with the expectation that data will likely be of higher quality from the controlled environment. The external evaluation dataset represents real-world use, where the signals are taken over longer periods of time and may include a higher degree of artifact due to uncontrollable environmental variation. We provide a comparison of both datasets in the supplement.

*Performance evaluation criteria*

DeepBeat takes pre-specified windowed physiological signal data of time length *t (25 seconds)* as input and performs two prediction tasks, signal quality assessment, and AF rhythm diagnosis. Figure 1 provides examples of physiological signals and quality assessment scores that were used to train and evaluate our method. We systematically compare the performance of the model on both a held-out test data and the external evaluation dataset. Table 1 reports the test performance of each algorithm explored. The performance of all models designed was evaluated based on four metrics: precision (the fraction of predicted diagnoses that match the expert diagnoses), recall (the fraction of the expert diagnoses that are successfully retrieved), F1 (the harmonic mean of precision and recall), and area under the precision-recall curve, aucPR, (high area under the curve represents both high recall and high precision). While accuracy is classically used for evaluating overall performance, we chose not to consider it here due to class imbalance, the F1 and auPRC are more appropriate metrics for determining rare AF detection rates.

*A feature-based approach has low discriminatory power for the detection of AF*

For a baseline comparison, we investigate the effectiveness of a traditional feature-based machine learning method against our model. Baseline results are reported in Table 1. Nine features most commonly used in feature-based AF detection and signal quality estimation algorithms[14–16] were calculated for input into a MultiOutputClassifier random forest model from scikit-learn[22]. Our model substantially outperforms the trained random forest model for AF detection across all metrics considered. The random forest results of precision:0.24, recall:0.58, f1:0.34, auPRC:0.44 was dramatically less effective for detecting AF events compared to DeepBeat's AF event metrics, precision:0.90, recall:0.98 f1:0.96, aucPR:0.96. Our model's notable improvement in AF detection over other a feature-based method is reflective across all DeepBeat versions examined, demonstrating that a feature-based approach fails to have high discriminatory power for AF detection.

*Mulitask learning is essential for high classification accuracy of AF observations.*

Training a single model on multiple tasks with shared encoding can improve a model's performance on all tasks, as different tasks serve as implicit regularization to prevent the model from overfitting to a particular task[23]. In our experiment, we investigated the performance of the model by considering the our mode's AF task individually as a single-task learner (STL) and then jointly with the additional signal quality assessment (QA) task. As shown in Table 2, we observe an improvement when considering the additional QA auxiliary task, F1-scores and aucPR raise from 0.41 and 0.48 to 0.64 and 0.68 respectively. The additional QA auxiliary task allows for signal QA thresholding to occur and removing false-positive AF detection events due to poor signal quality.

*Pretraining using CDAE increases classification accuracy.*

In order to investigate the effects of using CDAE as a pretraining method and examine the impact it might have on performance, we systematically compare the DeepBeat architecture with CDAE pretraining and without. Both model versions, Multi-task DeepBeat + pre_CDAE and Multi-task DeepBeat + no pretrain were trained under similar conditions and under the same parameters. In the experiments reported in Table 2, we find that using the extracted encoder from the trained CDAE as a form of unsupervised pre-training results in substantially higher performance across all performance metrics. The rates of precision dramatically increase from 0.56 to 0.94 and both the F1 score and auPRC increase from 0.72, 0.77 to 0.96 and 0.96 respectfully. These results show that the proposed Multi-task DeepBeat + pre_CDAE framework achieves high performance and using only a subset of these components leads to strictly worse test performance, demonstrating that they are all required to produce the optimal results.

*Interpreting model predictions*

In order to improve our understanding of how our model classifies AF events, we implemented a simple class activation map for visualization. Heatmaps of class activations over input signal were used to visualize how each data point within the signal influences the model's predictions. We show an example of a test example highlighted by saliency scores in Figure 3—the higher the saliency score, the lighter the color and the more influential is the area of the signal is to the model's prediction. In addition, we extract the output of the last dense layer before rhythm prediction is made and use UMAP[24] to visualize the learned distinction of non-AF versus AF event windows, Figure 4.

*DeepBeat achieves strong performance on a prospective ambulatory cohort*

Achieving high discriminatory power on a test dataset that originates from the same population distribution as the training dataset can be profound and significant. Transferring that same trained model to a different population and expecting high-performance metrics is usually unrealistic. In order to determine the robustness of our trained model, we test our model on a data from a prospective study of 15 free-living individuals. Data was collected over the course of two weeks using a different wrist-worn device. subjects simultaneously wore a rhythm patch device[25]. Rhythm was determined by clinician manual annotation following computerized reference ECG algorithms. Eleven of the 15 monitored individuals had no confirmed AF episodes during their two week period. We ran the DeepBeat AF event detection algorithm on these individuals to get an idea of the false positive rates this algorithm would have in an everyday setting. We partitioned the data from each individual into 25-second non-overlapping windows for AF classification and signal quality assessment and selected for excellent signal quality scored windows. Our results show less than a 0.01% false-positive detection rate across these individuals. The four individuals who had a confirmed AF event during the two weeks resulted in a total of 958 25-second AF classified episodes. DeepBeat classified AF presence in 3 out of 4 individuals (Table 3), detecting 925 episodes correctly (episode sensitivity of 0.97).

## Discussion

The primary contributions of this work are as follows:

1) We introduce a multi-task convolutional neural network method, DeepBeat, to model the intrinsic properties of physiological PPG signals. The proposed algorithm performs collaborative multi-task feature learning for two correlated tasks, input signal quality assessment and event detection (AF presence).

2) The model benefits from unsupervised transfer learning from pre-training using convolutional denoising autoencoders (CDAE) on simulated physiological signals. The ability of CDAEs to extract repeating patterns in the input makes them suitable to be used to extract true physiological signal from noise-induced inputs. Therefore, we use CDAE, for pretraining the foundational layers of the DeepBeat model.

3) Lastly, we explore the robustness of our model in a prospective external validation study in free-living ambulatory individuals monitored over a two-week time span. We find that DeepBeat maintains high discriminatory power in this cohort.

Reproducibility is a critical aspect of any clinical physiological measurement. In this work, we show the utility of scoring and incorporating quality signal assessment for event detection in a jointly trained deep learning approach. Incorporating a quality score allows for filtering out of unusable signal and assisted for *DeepBeat* achieved a high-performance metrics across precision, recall, F1-score, and auPRC on only 25 seconds sampling window. The high performance metrics could be attributed to two complementary components of the method: 1) unsupervised transfer learning: pretraining using CDAE on a simulated dataset and fine-tuning to learn features that were specific to the dataset, and 2) the use of a multi-task learning architecture, which had different capabilities in generalizing and adapting to different training objectives.

Previously, Poh et al[6] reported a method using dense convolutional neural networks (DCNN) with lower sensitivity (recall) of 95.2%, and PPV (precision) of 72.7% using single channel PPG channels as input to differentiate between AF and non-AF signals. This study included a large number of participants, but a very different methodological design. These authors consider noise as an equally likely category as AF and non- AF. The DCNN approach may limit the ability to disentangle the impact that signal quality has on predicted AF versus non-AF states. In addition, the DCNN architecture consisted of six dense blocks resulting in a model with a total of 201 layers, significantly deeper than that we propose here. Employing transfer learning with a much shallower network as seen with *DeepBeat* can increase the precision of AF events detected and an overly complicated model may not be necessary.

A substudy of the eHeart study evaluated the applicability of AF detection using a smartwatch[5]. The study design was broken into three distinct phases and the accuracy to detect AF was only moderate in the ambulatory stage. The performance metrics of sensitivity (0.98) and specificity (0.90) for the validation phase is comparable to the sensitivity reported here. The reduced specificity of their proposed model could be contributed to by the input data, use of averaged heart rates, step count, and time lapse. The ability to correctly identify average heart rate is critical for high-performance success and under varying conditions such as high-intensity motion or improper wear can greatly impact heart rate calculations. Out method is not based on calculated heart rate but instead infers rhythm directly from the raw waveform, reducing the need to calculate heart rate and reducing possible error propagation through the model. An additional benefit is that a prediction can be made instantly on just a brief window of rhythm monitoring.

A recent study, the Apple Heart Study, aimed for ambulatory AF detection with a proprietary algorithm based on irregular tachograms (periodic measurements of heart rate regularity). The study enrolled approximately 400,000 individuals for ambulatory AF monitoring, with 84% of irregular pulse notifications were concordant with atrial fibrillation[26]. The study was the largest to date and highlights the benefits of ambulatory monitoring in high risk populations.

A drawback of this framework includes examining other types of arrhythmias, we only considered one abnormal cardiac rhythm (albeit the most common one). We focused the limited collected training data to a mixture of healthy individuals and subjects who were hospitalized long term or for same day outpatient procedures (ie. taken from a population with a much higher prevalence of arrhythmia than the general population). When comparing reported evaluation metrics, our algorithm outperforms other deep learning methods that have been proposed for AF event detection [4-9]. However, a direct comparison of prior published work is challenging due to the lack of published code or released trained models and data for baseline neural network-based comparisons.

In summary, event detection of AF through the use of wearable devices is possible with strong diagnostic performance metrics. We achieve these through signal quality integration and data simulation and the use of CDAEs which are highly suitable for deriving features from noisy data. In light of the increased adoption of wearable devices and the need for cost-effective out of clinic patient monitoring, our method is well suited to serve as a method for extended rhythm monitoring in high-risk AF individuals or large-scale population AF screening. This study exemplifies how deep learning can be used to derive and detect digital event detection phenotypes and generate new insights into AF detection and disease progress in general. Wearable devices will allow for continuous AF detection that will likely provide personalized preventive health outcomes.

## METHODS

**Human Studies**

*Overview*

The source data used for training DeepBeat comprised a combination of a novel data generated for this study and publicly available data. Pre-training using CDAE was trained with a novel PPG simulated dataset, and DeepBeat was developed using subjects from two datasets from Stanford hospital, first subjects undergoing elective cardioversions and secondly, subjects performing elective stress tests. In addition, a publicly accessible 2015 IEEE Signal Processing Cup Dataset was used to supplement the Stanford dataset, to provide out of institution examples. For an additional evaluation, subjects from an external evaluation from an ambulatory cohort were used to evaluate algorithm performance. The subject demographics summary can be found in Table 1.

*Simulation of Synthetic Physiological Signals*

Simulated synthetic physiological PPG signals were generated, built upon RRest, a simulation framework [27]. The simulation framework for synthetic physiological signals was expanded to include a combination of baseline wander and amplitude modulation for simulation of sinus rhythm physiological signals. For simulations of an AF state, a combination of frequency modulation, baseline wander, and amplitude modulation was simulated. Frequency modulation was applied to specifically to mimic the chaotic irregularity of an AF rhythm. This assumption was the foundation for all simulations of AF signals. In addition to the expanded simulation version, an additional noise component was added to the simulated signals based on a Gaussian noise distribution. This provides the capability to simulate high-quality signals in the presence of low noise and low-quality signals in the presence of high noise. We simulated sinus rhythm and AF states under different levels of Gaussian noise to best represent observed real-world scenarios (Supplement).

*Collection of Subject Physiological Signals Before Cardioversion*

Physiological signals were derived from a wrist-based photoplethysmography wearable device worn by subjects at Stanford hospital undergoing direct current cardioversion for the treatment of AF underwent physiological monitoring with a wrist-based photoplethysmography wearable device. The study participants included subjects with a diagnosis of AF or atrial flutter (AFL) who were scheduled for elective cardioversion.  We included all adult subjects able to provide informed consent and willing to wear the device before and after the CV procedure. We included all subjects with an implanted pacemaker or defibrillator and who also had planned or unplanned transesophageal echocardiogram. In total 132 subjects were recruited and monitored for some period. Data from 107 subjects were of sufficient duration and quality to be included in this study. The average time for monitoring was approximately 20 minutes post and 20 minutes prior to the CV. All physiological signals were sampled at 128 Hz and wirelessly transmitted via Wifi to a cloud-based storage system..

*Collection of subject Physiological Signals from Exercise Stress Test*

Physiological signals were derived from a wrist-based photoplethysmography wearable device worn by subjects at Stanford hospital who were scheduled for an elective exercise stress test. We included all adult subjects who were able to provide informed consent and willing to wear the Simband during an elective exercise stress test. In total 42 subjects were monitored. Data from all 42 subjects were included in this study. The average time for monitoring the subject was approximately 45 minutes. All physiological signals were sampled at 128 Hz and wirelessly transmitted via Wifi to a cloud-based storage system.

*2015 IEEE Signal Processing Cup Dataset*

The PPG database from the 2015 IEEE Signal Processing Cup [28] was included in this study to provide a source of data from healthy participants. The dataset consists of two channels of PPG signals, three channels of simultaneous acceleration signals, and one channel of simultaneous ECG signal. PPG signals were recorded from a subject's wrist using PPG sensors built-in a wristband. The acceleration signals were recorded using a tri-axial accelerometer built into the wristband. The ECG signals were recorded using standard ECG sensors located on the chest of participants. All signals were sampled at 125 Hz and wirelessly transmitted via Bluetooth to a local computer.

*Prospective cohort*

Fifteen subjects with paroxysmal atrial fibrillation were recruited prospectively for a free-living ambulatory monitored over the course of two weeks, together with an ECG reference device. During the monitoring period, participants were asked to continue with their regular daily activities in their normal environment, and were asked to perform 8-10 ECG spot checks (short ECG recordings) and log times in which you performed those measurements. The dataset consists of a clinically-annotated dataset fully annotated by clinicians using a reference ECG device.

**Data Preprocessing**

Preprocessing of the simulated physiological signals for CDAE comprised of partitioning the data into training, validation and test partitions. Simulated physiological PPG signals consisted of 25 second time frames. The collected physiological signals were partitioned into training, validation and test partitions with no individual overlap between each set. We used overlapping windows for the training set as a data augmentation technique to increase the number of training examples. All signals were standardized to [0,1] bounds and bandpass filtered and downsampled by a factor of 4. Supplemental Table

S5 illustrates the number of signals for each partition from the Stanford cardioversion, exercise stress test and IEEE signal challenge datasets.

**Signal Quality Assessment Dataset**

In order to train a multi-task model assessing both signal quality and event detection, signal quality labels were needed for each signal window within the datasets. Event detection labels were known given the datasets and timestamp the signal originated from. To provide a signal quality assessment label for the training set we created an expert scored dataset of PPG signals known as the signal quality assessment dataset. For each time window set considered, 1,000 randomly selected windows were scored and partitioned into a train, validate and test sets. Each window was scored according to 1 of 3 categories (Excellent, Acceptable, Noise) in the concordance of published recommendations for PPG signal quality. The signal quality of signal data was based on standardized criteria (Elgendi quality assessments [17]). A separate model for QA was trained using the scored dataset as outcomes and used to predict quality labels for the remaining unscored windows considered.

**Algorithm**

*Unsupervised pre-training using* convolutional *denoising autoencoders*

Autoencoders are a type of neural network that is composed of two parts, an encoder, and decoder. Given a set of unlabeled training inputs, the encoder is trained to learn a compressed approximation for the identity function so that the decoder can produce output similar to that of the input, using backpropagation. Consider an input $x \in \Re^d$ being mapped to a hidden compressed representation $y \in \Re^d$ by the encoder function: $Encoder: y = h_\theta(x) = \sigma(Wx + b)$, where $W$ is the weights matrix, $b$ is the bias array, $\theta = \{W, b\}$ and $\sigma$ can be any nonlinear function such as ReLu. The latent representation $y$ is then mapped back into a reconstruction $z$, with the same shape as input $x$ using a similar mapping: $Decoder : z = \sigma(W'y + b')$. The reconstruction of the autoencoder attempts to learn the function such that $h_\theta(x) \approx x$, to minimize the mean squared difference $L(x, z) = \sum(x - h_\theta(x))^2$.

Convolutional denoising autoencoders (CDAE) are a stochastic extension to traditional autoencoders explained above. In CDAE, the initial input $x$ is corrupted to $\bar{x}$ by a stochastic mapping $\bar{x} = C(\bar{x}|x)$, where C is a noise generating function, which partially destroys the input data. The hidden representation $y$ of the $k$th feature map is represented by $y^k = \sigma(W^k * x + b)$, where $*$ denotes the 1D convolutional operation and $\sigma$ can be any nonlinear function. The decoder is denoted by

$z \approx h_\theta(x) = \sigma(\sum_{i \in m} m^i * \overline{W} + b)$, where $m$ indicates the group of latent feature maps and $\overline{W}$ is the flipped operation over the dimensions of $W$ [29]. Compared to traditional autoencoders, convolutional autoencoders can utilize the full capability of convolutional neural networks to exploit structure within the input with weights shared among all input locations to help preserve local spatiality [29].

CDAE have been applied for unsupervised pre-training [23] and can be categorized as a data-driven network initialization method or a special type of semi-supervised learning approach[30]. We simulated a training dataset for artifact induced PPG signals and its corresponding clean/target signal. We use convolutional and pooling layers in the encoder, and upsampling and convolutional layers in the decoder. To obtain the optimal weights for $W$, weights were randomly initiated according to He distribution [31] and the gradient calculated by using the chain rule to back-propagate error derivatives through the decoder network and then the encoder network. Using a number of hidden units lower than the inputs forces the autoencoder to learn a compressed approximation. The loss function employed in pretraining was mean squared error (MES) and was optimized using a back-propagation algorithm. The input to the CDAE was the simulated signal dataset with a Gaussian noise factor of 0.001, 0.5, 0.25, 0.75, 1, 2, and 5 added to corrupt the simulated signals. The uncorrupted simulated signals are than used as the target for reconstruction. We used three convolution layers and three pooling layers for the encoder segment and three convolution layers and three upsampling layers for the decoder segment of the CDAE. Further details of model architecture can be found in the Supplement. ReLU was applied as the activation function and Adam [32] is used as the optimization method. Each model was trained with mean squared error (MSE) loss for 200 epochs with a reduction in learning rate by 0.001 for every 25 epochs if validation loss did not improve. Further results from the CDAE training can be found in the supplemental materials.

*Transfer learning and fine-tuning of multi-task DeepBeat Model*

Transfer learning is an appealing approach for problems where labeled data is acutely scarce[33]. In general terms, transfer learning refers to the process of first training a base network on a source dataset and task before transferring the learned features (the network's weights) to a second network which is trained on an external and sometimes related dataset and task. The power of transfer learning is rooted in its ability to deal with domain mismatch. Fine-tuning pre-trained weights on the new dataset is implemented by continuing backpropagation. It has been shown that transfer learning reduces the training time by reducing the number of epochs needed for the network to converge on the training set [34]. We utilize transfer learning here by extracting the encoder weights from the pre-trained CDAE and copy the weights to the first 3 layers of the *DeepBeat* model architecture. The motivation behind using CDAE for unsupervised pre-training on simulated physiological signals was to provide the earlier foundational layers of the *DeepBeat* model the ability to quickly identify learned features that constitute important physiological signal elements.

*DeepBeat Network Structure*

Following the success of other multi-task learning neural networks [35], the *DeepBeat* network structure for AF event detection and signal quality assessment was designed to leverage several advantages that multi-task learning neural networks possess.

The *DeepBeat* architecture utilizes low-level feature sharing by allowing the signal quality assessment task and AF event detection task to initially share the same learned feature representation through the use of shared layers or hard parameter sharing. This is motivated by the following: First, learned features for the signal quality assessment task provide value for analyzing regions for which AF events are present or absent. Second, shared layers encourage a reduction in the number of parameters needed for the network to generalize on a greater range of individuals. The integration of signal context information is beneficial. Difficulties arise when considering AF events and non-AF event signals that are corrupted by noise or artifact. One approach to overcoming these obstacles involves the provision of context information regarding signal quality so the learned features are in tandem with features predictive of AF events. The event detection task determines whether a window contains regions of AF events, while the signal assessment task predicts the quality of the signal. Preservation of important physiological signal information throughout the model should not be translation invariant, i.e., the learned features from the signal assessment task should be preserved in the AF detection outputs.

*DeepBeat* was trained to classify two tasks through an initially shared structure. The input is a single physiological PPG signal. The first three hidden layers are the extracted encoder architecture from the CDAE pre-trained on the separate simulated physiological signal dataset. In the first three shared hidden layers, convolutional layers and pooling layers are used to ensure the activation of neurons are affected by only local patterns from the input signal. Rectified Linear Unit (ReLU), and max-pooling layers were added after each convolutional layer to increase non-linearity and to consolidate local information. In addition, small filter and stride sizes were selected to control for localized signal information. The convolutional layers of the first three layers include receptive field maps with initialized filter weights from the pre-trained CDAE encoder section. Three additional layers were added after the encoder section, leading to a total of 6 shared hidden layers. For hidden layers 4-6, leaky rectified linear unit [36], batch normalization [37], and dropout layers and convolutional parameters were selected through hyperparameter search.

The Quality Assessment task (QA) and event detection task builds on upon the six shared layers, before branching into two specialized branches, the QA task and event detection task. The QA branch is comprised of an additional convolutional layer, Rectified Linear Unit (ReLU), batch normalization, dropout and two dense layers before final softmax activation for classification was used. The event detection task consisted of three additional convolutional layers, each followed by a Rectified Linear Unit (ReLU), batch normalization, and dropout layers. Two additional dense layers were also added before final softmax activation for event detection. The *DeepBeat* model was trained with a category cross-entropy loss function and hyperparameter optimization for the number of layers, activation functions, receptive field map size, convolutional filter length, and stride, using hyperas [38]. With a given training input, predictions for both tasks, QA and rhythm event detection were calculated and backpropagation was used to update the weights and the corresponding gradients throughout the network. We implemented *DeepBeat* using Python 3.5, Keras [39] with Tensorflow 1.2 [40]. We trained the model on a cluster of 4 nodes each having 25GB memory and accelerated by an NVIDIA P100 GPU.

**DeepBeat performance metrics**

To evaluate the performance of *DeepBeat* on the collected set of real-world data, the data was split into training, validation, and test set with no subject overlap between sets, as described in Methods. Classification performance was measured by using the following four metrics: precision, recall, F1 score (F1), and Area under the Precision-Recall Curve (AuPRC). While accuracy is classically used for evaluating overall performance, F1 and auPRC are more useful with significant class imbalance, as is the case here.

## Data availability

The data used to train DeepBeat will be available upon request and algorithm and code details can be found at: https://github.com/AshleyLab/deepbeat

## Acknowledgements

We thank our academic and technology partners who helped with the project, with special thanks to Jeff Christle and Samsung partners at Samsung Strategy and Innovation Centre.

## Author Contributions

JTS, and EAA conceptualised and designed the study. JTS designed and coded algorithm. JTS, and EAA drafted and critically revised the manuscript for important intellectual content.

## Competing Interests statement

EAA reports advisory board fees from Apple, DeepCell, Myokardia, and Personalis, outside the submitted work. All other authors declare no competing interests.

# FIGURES

**Figure 1. Example of rhythm and quality assessments of training data.**
Examples of physiological signals grouped by assessment scores that was used to train and evaluate the Deepbeat model. From top to bottom, quality assessment score excellent, acceptable and poor. Left column are signals from subjects in normal sinus rhythm and right column are signals from subjects in atrial fibrillation.

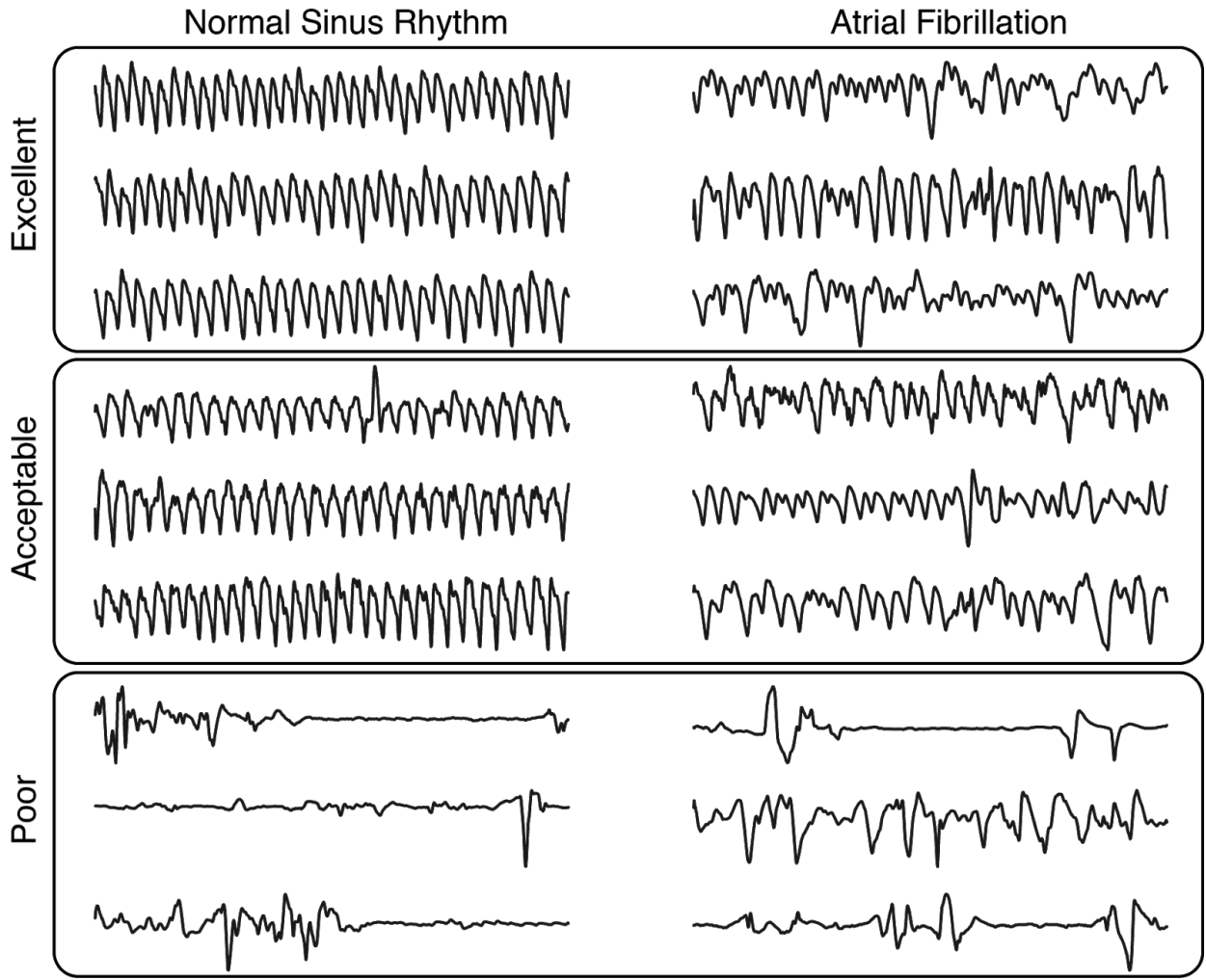

**Figure 2. Deepbeat model architecture.**

The proposed model architecture for Deepbeat, two tasks are shown: (top) unsupervised pre-training and (bottom) supervised learning through fine tuning. The top represents the pre-training process on the unlabeled simulated data, and the bottom represent the multi-task fine-tuning process on the labeled data. The trained encoder weights serve as the foundational layers of the multi-task model.

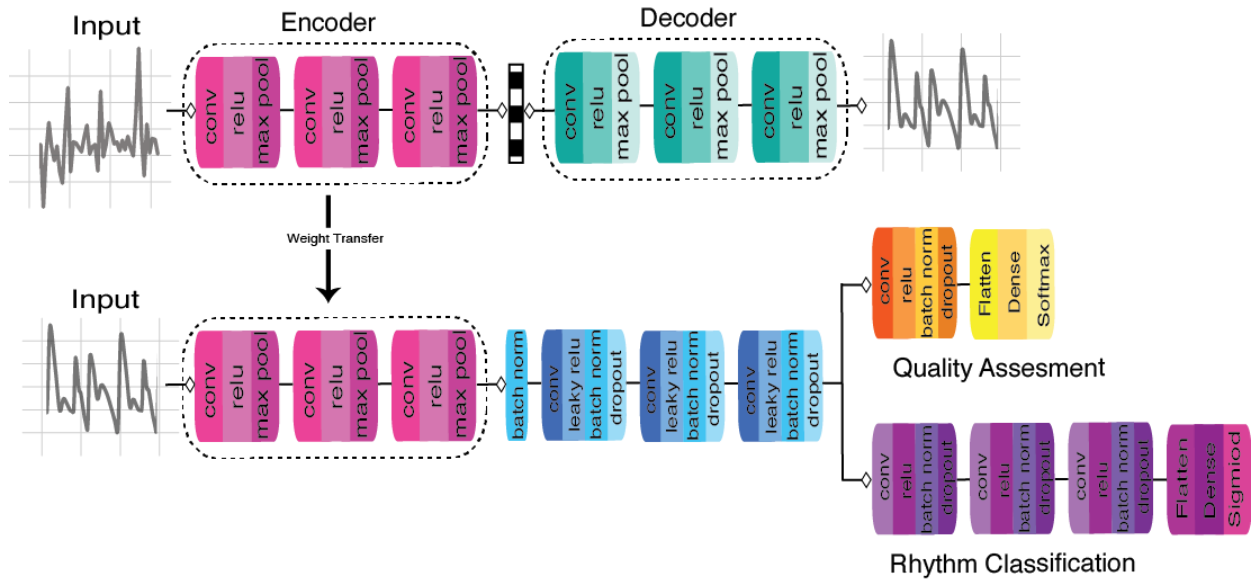

**Figure 3. Differences in class activation map by rhythm classification.**

Example of signals from held out test dataset. The predicted class score is mapped back to the last convolutional layer to generate the class activation maps (CAMs). The CAM highlights the class-specific discriminative regions between sinus (top) and atrial fibrillation (bottom).

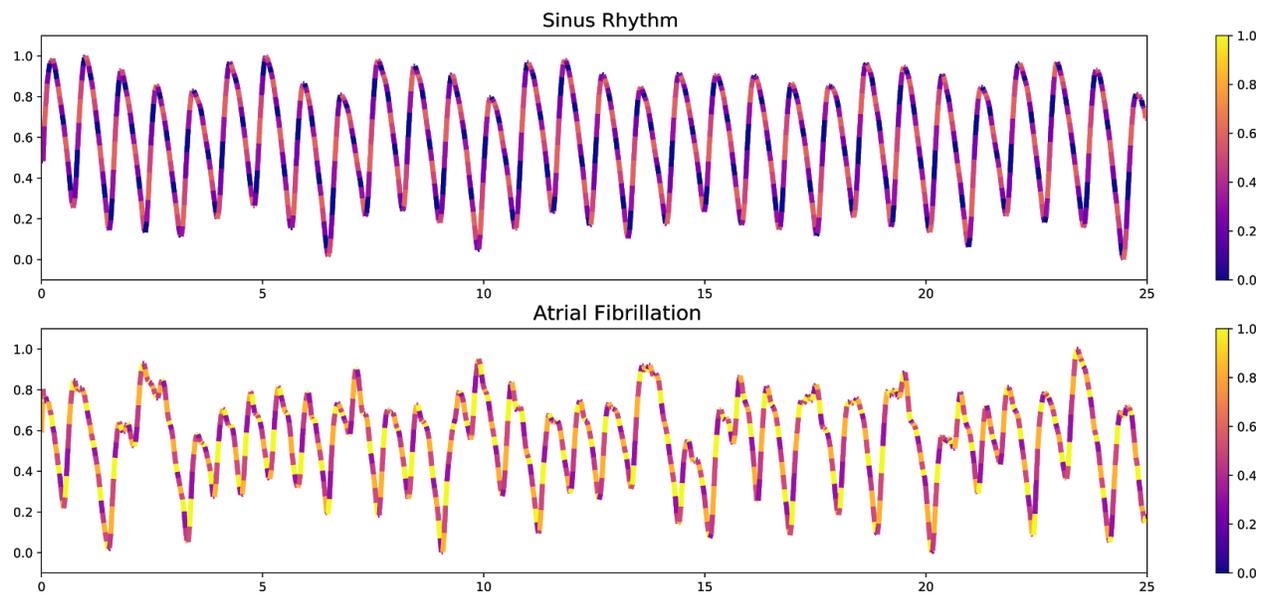

**Figure 4. Visualization of learned rhythm class distinction.**
UMAP representation at the last (Dense_18) layer of the DeepBeat model. The two colors represent the two classes, respectively normal sinus rhythm as purple, and yellow as atrial fibrillation.

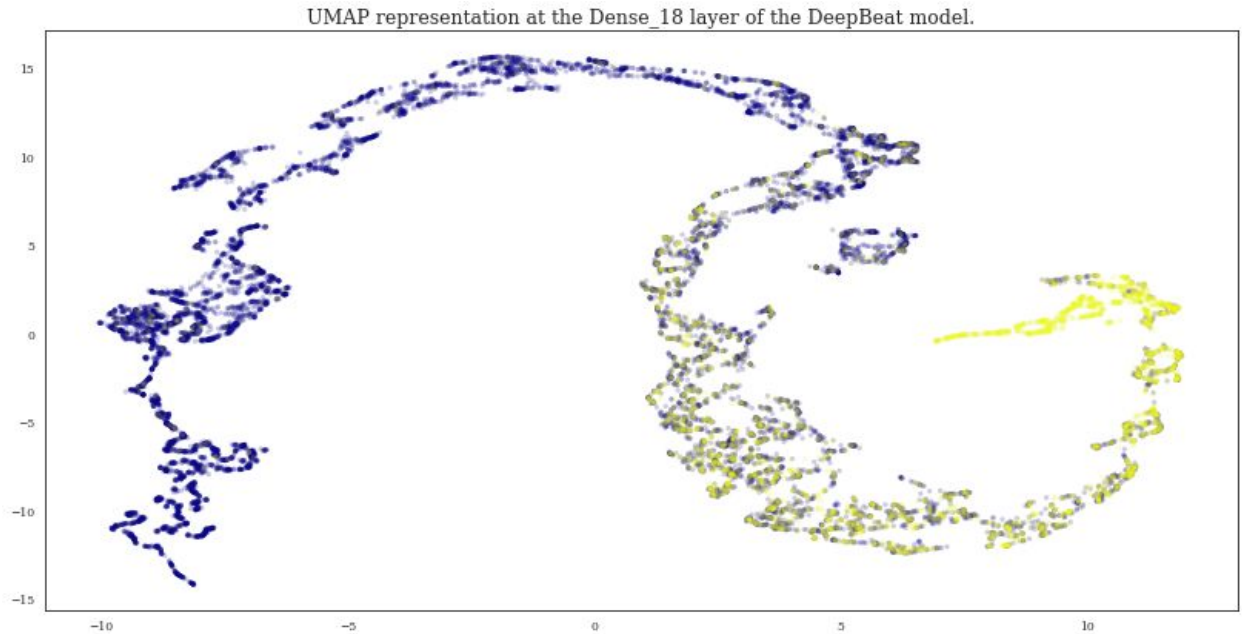

# TABLES

**TABLE 1** Demographics of study subjects used for development of DeepBeat

|  | Cardioversion cohort | Exercise stress test cohort | IEEE Signal Processing Cup[1] | Ambulatory cohort |
|---|---|---|---|---|
| Number of subjects | 107 | 41 | 20 | 15 |
| Number of subjects with Atrial Fibrillation | 107 | 0 | 0 | 5 |
| Mean age | 68 | 56 | 26 | 67 |
| Sex (M/F) | 85/22 | 27/14 | 7/1 | 11/4 |

[1] Demographics from held out test set only

**TABLE 2** Evaluation of trained classifiers on held-out test data set

| Model | Precision | Recall | F1-score | auPRC |
|---|---|---|---|---|
| Random Forest | 0.24 | 0.58 | 0.34 | 0.44 |
| DeepBeat:Single task (no pretrain CDAE + AF) | 0.42 | 0.41 | 0.41 | 0.48 |
| DeepBeat:Single task (pretrain CDAE + AF) | 0.59 | 0.69 | 0.64 | 0.68 |
| DeepBeat:Multi-task (no pretrain CDAE+ AF + Excellent QA) | 0.56 | 0.98 | 0.72 | 0.77 |
| **DeepBeat:Multi-task (pretrain CDAE+ AF + Excellent QA)** | **0.94** | **0.98** | **0.96** | **0.96** |

**TABLE 3** Prospective evaluation in an ambulatory cohort

| Model | Precision | Recall | F1-score | auPRC |
|---|---|---|---|---|
| DeepBeat:Multi-task (pretrain CDAE+ AF + Excellent QA) | 0.94 | 0.97 | 0.94 | 0.94 |

## Supplemental Materials

*Results from unsupervised pre-training using* convolutional *denoising autoencoders*

Results from the trained CDAE on scored signal quality assessment dataset can be found in supplemental table S2. The mean squared errors were 0.0095, 0.0104, 0.0.0143 for excellent, acceptable and poor categories for the 25 second time segments. We found the lowest mean squared error for signal reconstruction in the excellent category across all time segments, suggesting that the trained CDAE is selecting filters appropriate for high-quality physiological signal reconstruction.

In order to determine that the CDAE was not introducing modulations typical of physiological signals when there was no physiological signal present, we performed a sensitivity analysis. Five hundred random signals were generated and run through the trained CDAE model. The estimated MSE of the randomly generated noise was similar to that of the estimated MSE for the poor signal quality category across all time points. To further explore the estimated reconstruction predictions from the output of the trained CDAE, predictions were compared to 3rd order Savitzky-Golay filters. Mean squared error of the reconstruction CDAE prediction of the randomly generated noise set was 0.026, and mean squared error of the 3rd order Savitzky-Golay filters was 0.023. These results confirm that the trained CDAE model provides a set of filters sensitive to frequencies unique to physiological signals and, in situations where no viable physiological signal is present, CDAE instead acts as a smoothing filter supplementary Figure S1.

*Details regarding random forest*

To investigate the choice of a multi-task model, a comparison of different methods was performed. For a feature-based approach, random forests were used due to its capability of finding complex nonlinear relationships in data. The following features were calculated: kurtosis, skew, entropy, zero crossings, hjorthe mobility, hjorthe complexity, normalized root mean of successive differences, and Shannon entropy. A MultiOutputClassifier random forest model with n_estimators=100 and random_state=1 was used as parameters for training.

## Supplemental Figures and Tables

Figure S1: Results from trained CDAE on denoising simulated signals, collected signal and random noise.

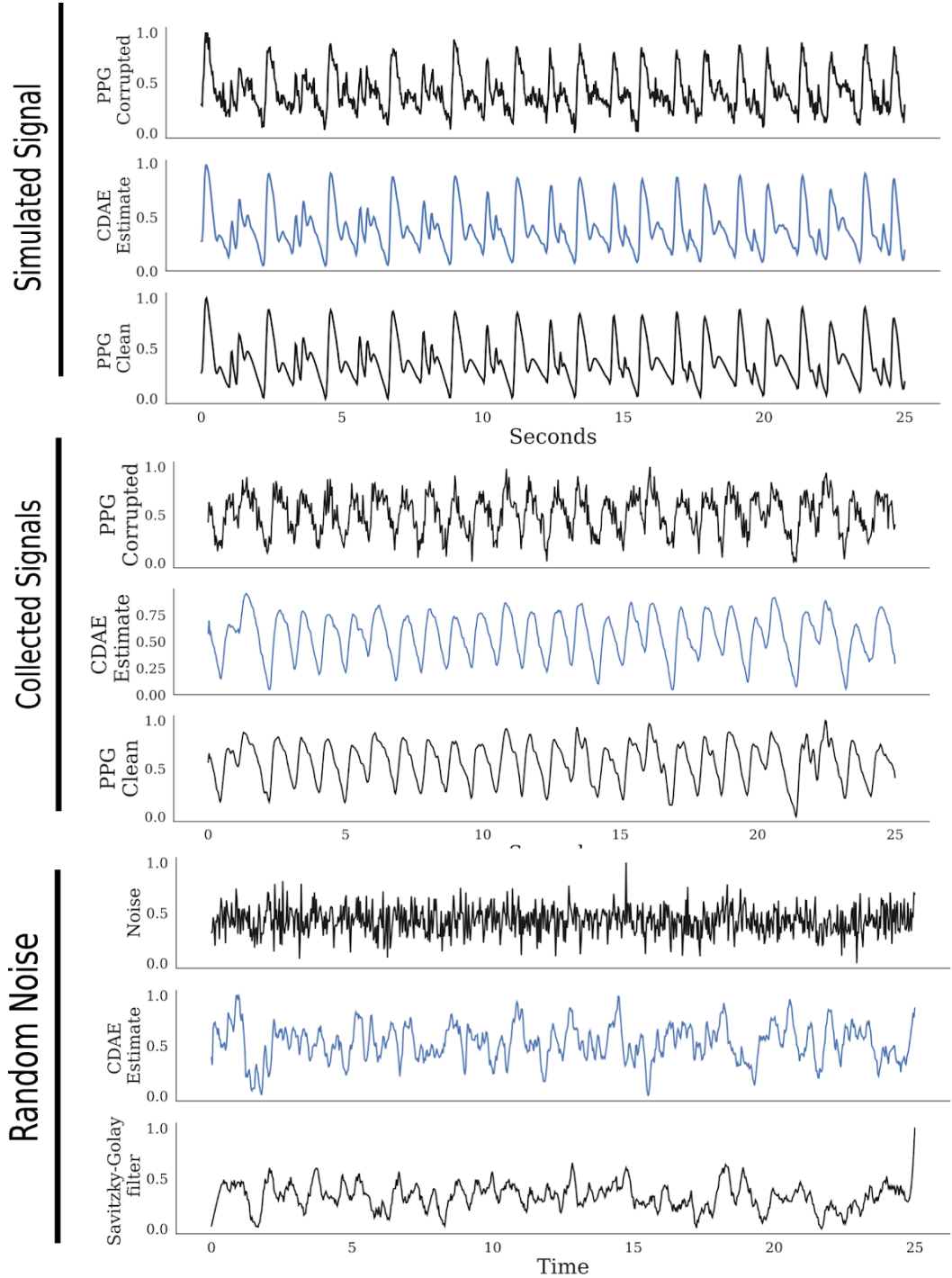

Figure S2 : Simulated signals from dataset A. Left column, simulated sinus rhythm, top to bottom in increasing order of added Gaussian noise mixture (0.001, 0.15, 0.5, 0.75, 1, 2, 5). Right column, simulated AF rhythm, top to bottom in increasing added Gaussian noise mixture (0.001, 0.15, 0.5, 0.75, 1, 2, 5). Pictured below all signals were simulated at 60 beats per minute (BPM).

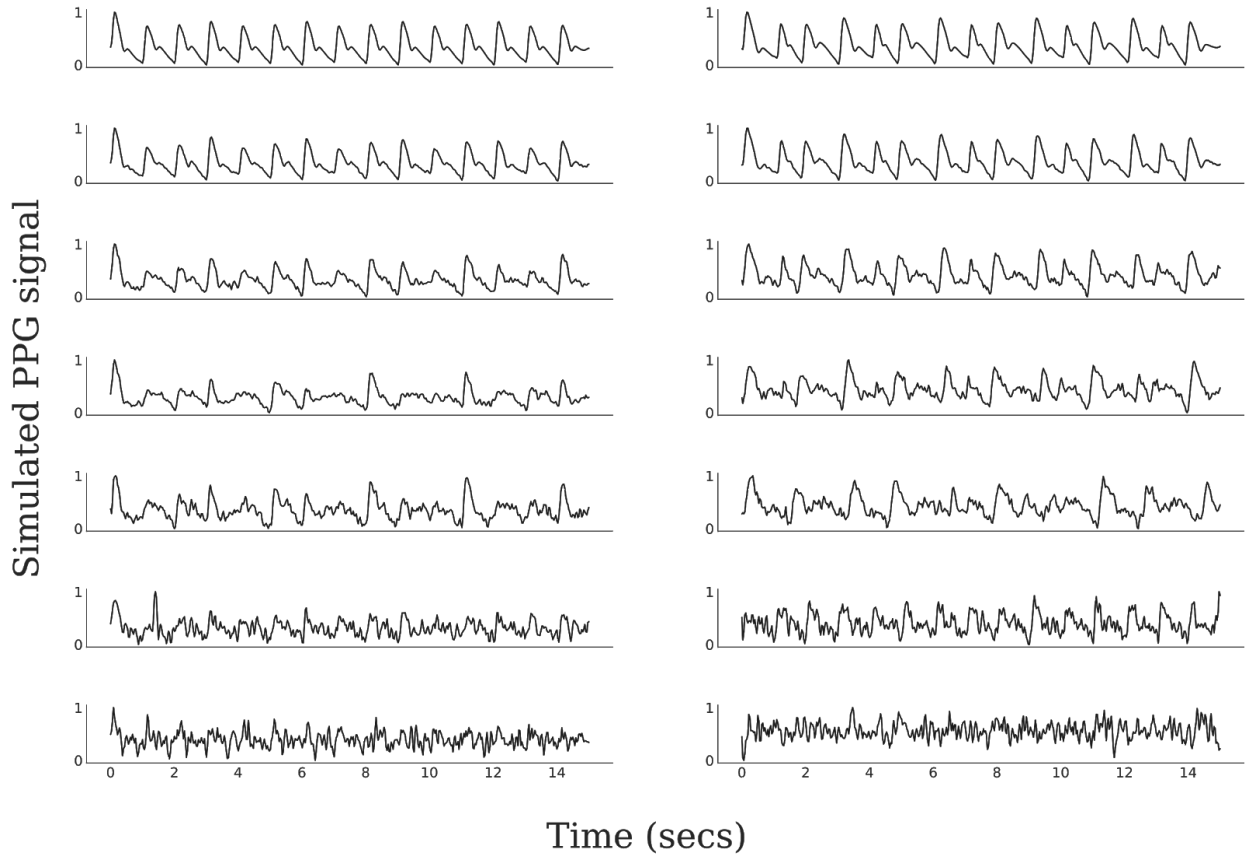

**TABLE S1** CDAE model architecture specifications

| Layer Type | Output Shape | Param # |
|---|---|---|
| **Encoder** | | |
| InputLayer | (None, 800, 1) | 0 |
| Conv1D | (None, 800, 64) | 704 |
| MaxPooling | (None, 266, 64) | 0 |
| Conv1D | (None, 266, 45) | 23,085 |
| MaxPooling | (None, 88, 45) | 0 |
| Conv1D | (None, 88, 50) | 11,300 |
| MaxPooling | (None, 44, 50) | 0 |
| **Decoder** | | |
| Conv1D | (None, 44, 50) | 12,550 |
| UpSampling | (None, 88, 50) | 0 |
| Conv1D | (None, 88, 45) | 18,045 |
| UpSampling | (None, 264, 45) | 0 |
| Conv1D | (None, 264, 64) | 28,864 |
| UpSampling | (None, 792, 64) | 0 |
| Flatten | (None, 50688) | 0 |
| Dense | (None, 800) | 40,551,200 |

**TABLE S2** CDAE MSE for Signal Reconstruction

| | Excellent | Acceptable | Poor |
|---|---|---|---|
| **25 seconds** | 0.0095 | 0.0104 | 0.0143 |

**TABLE S3** DeepBeat model architecture specifications

| Rhythm Branch | | | Layer Type | Output Shape | Param # | Quality Assessment Branch | | |
|---|---|---|---|---|---|---|---|---|
| | | | **Extracted Encoder** | | | | | |
| | | | InputLayer | (None, 800, 1) | 0 | | | |
| | | | Conv1D | (None, 800, 64) | 704 | | | |
| | | | MaxPooling | (None, 266, 64) | 0 | | | |
| | | | Conv1D | (None, 266, 45) | 23,085 | | | |
| | | | MaxPooling | (None, 88, 45) | 0 | | | |
| | | | Conv1D | (None, 88, 50) | 11,300 | | | |
| | | | MaxPooling | (None, 44, 50) | 0 | | | |
| | | | **Shared layers** | | | | | |
| | | | BatchNormalization | (None, 44, 50) | 200 | | | |
| | | | Conv1D | (None, 15, 64) | 12864 | | | |
| | | | Leaky ReLu | (None, 15, 64) | 0 | | | |
| | | | BatchNormalization | (None, 15, 64) | 256 | | | |
| | | | Dropout | (None, 15, 64) | 0 | | | |
| | | | Conv1D | (None, 5, 35) | 8995 | | | |
| | | | Leaky ReLu | (None, 5, 35) | 0 | | | |
| | | | BatchNormalization | (None, 5, 35) | 140 | | | |
| | | | Dropout | (None, 5, 35) | 0 | | | |
| | | | Conv1D | (None, 5, 64) | 9024 | | | |
| | | | Leaky ReLu | (None, 5, 64) | 0 | | | |
| | | | BatchNormalization | (None, 5, 64) | 256 | | | |
| Conv1D | (None, 2, 35) | 11235 | Dropout | (None, 5, 64) | 0 | Conv1D | (None, 3, 25) | 6425 |
| BatchNormalization | (None, 2, 35) | 140 | | | | BatchNormalization | (None, 3, 25) | 100 |
| Dropout | (None, 2, 35) | 0 | | | | Dropout | (None, 3, 25) | 0 |
| Conv1D | (None, 1, 25) | 525 | | | | Flatten | (None, 75) | 0 |
| BatchNormalization | (None, 1, 25) | 100 | | | | Dense | (None, 175) | 13300 |
| Dropout | (None, 1, 25) | | | | | Dense | (None, 3) | 528 |
| Conv1D | (None, 1, 35) | 2660 | | | | | | |
| BatchNormalization | (None, 1, 35) | 140 | | | | | | |
| Dropout | (None, 1, 35) | 0 | | | | | | |
| Flatten | (None, 35) | 0 | | | | | | |
| Dense | (None, 175) | 6300 | | | | | | |
| Dense | (None, 2) | 352 | | | | | | |

**TABLE S4** Comparison of evaluation data sets

|  | Number of patients | Total number of windows | Total number of AF windows |
|---|---|---|---|
| Held out test set | 22 | 17,617 | 4,230 |
| Ambulatory cohort | 15 | 20,492 | 2,048 |

**TABLE S5** Window counts used for the DeepBeat model

| partition | rhythm | QA | Count |
|---|---|---|---|
| train | sinus | poor | 999,253 |
|  |  | acceptable | 140,020 |
|  |  | excellent | 390,851 |
|  | AF | poor | 812,237 |
|  |  | acceptable | 110,740 |
|  |  | excellent | 125,107 |
| validate | sinus | poor | 295,449 |
|  |  | acceptable | 56,572 |
|  |  | excellent | 119,154 |
|  | AF | poor | 33,691 |
|  |  | acceptable | 8,075 |
|  |  | excellent | 5,841 |
| test | sinus | poor | 8,856 |
|  |  | acceptable | 1,718 |
|  |  | excellent | 2,813 |
|  | AF | poor | 3,483 |
|  |  | acceptable | 314 |
|  |  | excellent | 433 |